\documentclass[aps,twocolumn,preprintnumbers]{revtex4}

\usepackage{graphicx}  
\usepackage{multirow}
\usepackage{float}

\usepackage{fancyhdr}
\usepackage{longtable}
\usepackage{parskip}
\usepackage[T1]{fontenc}
\usepackage{dcolumn}   

\usepackage{bm}        
\usepackage{amsfonts}  
\usepackage{amsmath}   
\usepackage{amssymb}   
\usepackage{siunitx}

\newcommand{\bra}[1]{\left\langle #1 \right|}
\newcommand{\ket}[1]{\left| #1 \right\rangle}

\begin{document}

\title{Valley separation of photoexcited carriers in bilayer graphene}

\author{T. J. Osborne}
\email{to45@st-andrews.ac.uk}
\altaffiliation{%
 Present address: School of Physics and Astronomy, University of St Andrews, North Haugh, St Andrews KY16 9SS, United Kingdom
}%
\affiliation{%
 Physics and Astronomy, University of Exeter, Stocker Road, Exeter EX4 4QL, United Kingdom
}%

\author{M. E. Portnoi}
\email{m.e.portnoi@exeter.ac.uk}
\affiliation{%
 Physics and Astronomy, University of Exeter, Stocker Road, Exeter EX4 4QL, United Kingdom
}%

\author{E. Mariani}
\email{e.mariani@exeter.ac.uk}
\affiliation{%
 Physics and Astronomy, University of Exeter, Stocker Road, Exeter EX4 4QL, United Kingdom
}%

\date{\today}%


\begin{abstract}  

We derive the angular generation density of photoexcited carriers in gapless and gapped Bernal bilayer graphene. Exploiting the strong anisotropy of the band structure of bilayer graphene at low energies due to trigonal warping, we show that charge carriers belonging to different valleys propagate to different sides of the light spot upon photoexcitation. Importantly, in this low-energy regime, inter-valley electron-phonon scattering is suppressed, thereby protecting the valley index. This \emph{optically induced} valley polarization can be further enhanced via momentum alignment associated with linearly-polarized light. We then consider gapped bilayer graphene (for example with the gap induced by external top- and back-gates) and show that it exhibits valley-dependent optical selection rules with circularly-polarized light analogous to other gapped Dirac materials, such as transition metal dichalcogenides. Consequently, gapped bilayer graphene can be exploited to \emph{optically detect} valley polarization. Thus, we predict an optical valley Hall effect - the emission of two different circular polarizations from different sides of the light spot, upon linearly-polarized excitation. We also propose two realistic experimental setups in gapless and gapped bilayer graphene as a basis for novel optovalleytronic devices operating in the elusive terahertz regime.

\end{abstract}

\maketitle 

\section{Introduction}

Valleytronics is an emerging technology that seeks to exploit the local extrema (known as valleys) in the electronic band structure of materials for the storing and processing of quantum information \cite{schaibley2016valleytronics,vitale2018valleytronics}. Many crystalline solids exhibit energy-degenerate valleys in their band structure. Notably, E. I. Rashba, one of the founding fathers of spintronics, made an early contribution to what would now be recognized as valleytronics more than six decades ago, by proposing a method for controlling the valley population in multivalley semiconductors using an electric field \cite{rashba1965redistribution}. However, selectively addressing these valleys in conventional semiconductors is usually very difficult, making such materials impractical for valleytronic devices.

Since the exfoliation of graphene in 2004, investigation into two-dimensional (2D) materials has shown that they are promising candidates for valleytronics due to their often strong valley-dependent interactions with applied electric and magnetic fields. Indeed, graphene has an electronic band structure characterized by two nonequivalent massless Dirac cones (valleys) at the corners of the Brillouin zone \cite{katsnelson2020physics}. The well defined valley quantum number of electrons in graphene spurred research into its feasibility for valleytronics. The first proposals for valleytronics applications in graphene were confined to the electronic transport regime. It was soon realized that atomic scale defects at the edges of realistic devices mix the valleys, hampering the potential for valleytronics in electronic transport. 

An alternative path to valleytronics involves selective optical excitations of charge carriers in the bulk of 2D materials. However, the inversion symmetry of pristine graphene results in valley-independent optical selection rules for interband transitions at low energy, posing a challenge for the creation and measurement of valley polarization. At higher excitation energies, when trigonal warping in the graphene spectrum becomes important, charge carriers in different valleys can be spatially separated in the instance of photocreation \cite{hartmannthesis,saroka2022momentum}. However, this anisotropy in the band structure occurs well above the energy of $0.16\ \si{eV}$ associated with ultra-fast inter-valley electron-phonon scattering \cite{phononscattering}, that dramatically reduces the lifetime of the valley quantum number and makes the effect impractical for quantum valleytronics.

The inversion symmetry of graphene may be broken by placing the sample on a matching substrate which makes the two sublattice sites nonequivalent, such as boron nitride \cite{boron}. This opens a band gap in the electronic dispersion resulting in strong valley-dependent optical selection rules. Electrons in one valley may be selectively excited by circularly-polarized light of a given handedness \cite{saroka2022momentum}. This effect is well-known in other 2D gapped Dirac materials, such as transition metal dichalcogenides (TMDs) \cite{schaibley2016valleytronics}. Valley population in gapped Dirac materials may indeed be controlled by the degree of circular polarization, however, purely optical spatial separation of charge carriers belonging to different valleys is not possible in the absence of warping. Moreover, opening the gap in Dirac materials results in low carrier mobility that affects the ability to propagate the valley index \cite{boron2, TMD}. 

In this paper, we propose a method to optically induce and detect sustainable spatial separation of charge carriers from different valleys in a high mobility, gapless Dirac material: Bernal-stacked bilayer graphene. There is currently a resurgence of interest in this material, with recent research revealing a variety of new effects, including many-body phase transitions and superconductivity \cite{bilayersuperconductivity,bilayercorrelated,bilayermanybody}. We consider low-frequency photoexcitation in bilayer graphene that has been overlooked so far.  In contrast to monolayer graphene, the valleys of bilayer graphene are highly anisotropic at the $\si{meV}$ scale. Firstly, we show that this anisotropy can be exploited to \emph{induce} spatial separation of charge carriers belonging to different valleys by illuminating the sample with low-frequency photons. This valley polarization may be further enhanced by exploiting momentum alignment with linearly-polarized light. In stark contrast to monolayer graphene, this effect occurs at low excitation energies where inter-valley phonon scattering is suppressed, thereby preserving the topological protection of the valley index in view of novel valleytronic applications. We subsequently demonstrate that the valley-dependent spatial separation of charge carriers persists even after opening a moderate band gap in the electronic dispersion.
Finally, we show that gapped bilayer graphene exhibits valley-dependent optical selection rules with circularly polarized light similar to those of other gapped 2D Dirac materials. These selection rules can be exploited to \emph{detect} the degree of valley polarization induced after the spatial separation of charge carriers.

By combining these phenomena, we propose two experimental setups to optically induce and detect valley separation in conventional bilayer graphene devices routinely available in  experimental labs worldwide.

\section{Bilayer Graphene}

\begin{figure}
\includegraphics[height = 160pt]{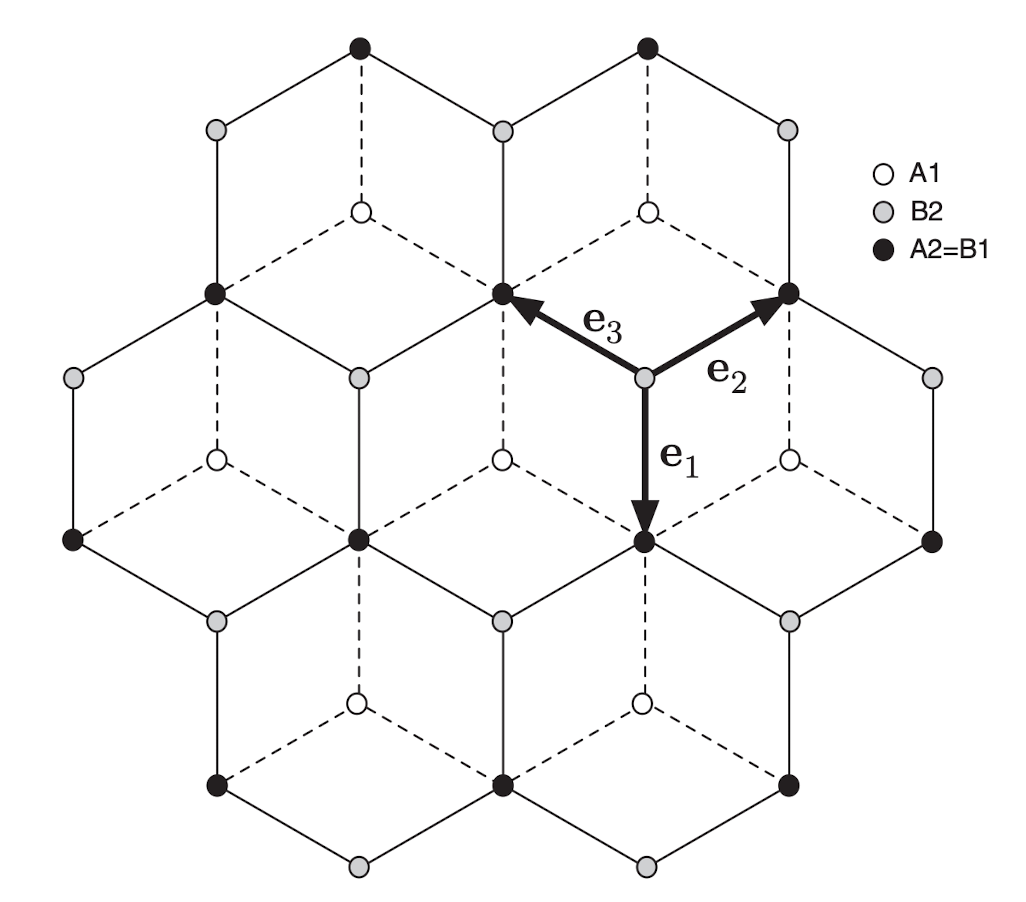}
\caption{Atomic structure of perfect bilayer graphene in Bernal stacking (top view). The honeycomb lattice with solid (dashed) lines corresponds to the upper (lower) layer. The three vectors $\bm{e}_j$, (j = 1, 2, 3) connecting A and B sites are indicated \cite{Mariani2012}.}
\label{fig:bilayer_lattice}
\end{figure}

The crystal structure of Bernal-stacked bilayer graphene is shown in Fig. \ref{fig:bilayer_lattice}. The second layer is shifted with respect to the first so that the atoms of sublattice A in layer 2 (A2) are located directly above atoms of sublattice B in layer 1 (B1), at a distance $c\simeq 3.34\ \text{\AA}$. The pairs A1-A2, B1-B2 and A1-B2 are all separated by the distance $\tilde{c}=\sqrt{c^2+a^2}\simeq 3.63 \ \text{\AA}$ where $a=1.42\ \text{\AA}$ is the nearest-neighbor distance in monolayer graphene \cite{Mariani2012}.

The electronic properties of bilayer graphene can be analyzed using the tight-binding Hamiltonian
\begin{equation}
    H=
    \begin{pmatrix}
        0 & -\gamma_3 f(\bm{k}) & 0 & -\gamma_0 f^*(\bm{k}) \\ -\gamma_3 f^*(\bm{k}) & 0 & -\gamma_0 f(\bm{k}) & 0 \\ 0 & -\gamma_0 f^*(\bm{k}) & 0 & -\gamma_1 \\ -\gamma_0 f(\bm{k}) & 0 & -\gamma_1 & 0
    \end{pmatrix},
    \label{eq:bilayerHamiltonian}
\end{equation}
keeping only the dominant terms relevant for describing the low-energy sector. The basis of electron states is $(\psi_{A1},\psi_{B2},\psi_{A2},\psi_{B1})^T$, where $\psi_{sl}$ is the amplitude of the wavefunction on the sublattice $s$ and layer $l$. The term $f(\bm{k})=\sum_{j} \exp{(i \bm{k} \cdot \bm{e}_j)}$ for the wavevector $\bm{k}$, with the nearest-neighbor vectors $\bm{e}_1=a(0,-1)^T$, $\bm{e}_2=a(\sqrt{3}/2,1/2)^T$, and $\bm{e}_3=a(-\sqrt{3}/2,1/2)^T$, as shown in Fig. \ref{fig:bilayer_lattice}. The hopping terms are $\gamma_0\simeq3.16 \si{eV}$, $\gamma_1\simeq0.38 \si{eV}$, and $\gamma_3\simeq0.38 \si{eV}$ \cite{mccann2013electronic}.

In the vicinity of the two inequivalent Dirac points (or valleys) $\bm{K_{\pm}}=\pm(4\pi/3\sqrt{3}a,0)^T$, the Hamiltonian in 
Eq.~(\ref{eq:bilayerHamiltonian}) can be expanded as
\begin{equation}
    H^{(+)}=
    \begin{pmatrix}
        0 & v_3 p & 0 & v_0 p^\dagger \\ v_3 p^\dagger & 0 & v_0 p & 0 \\ 0 & v_0 p^\dagger & 0 & -\gamma_1 \\ v_0 p & 0 & -\gamma_1 & 0
    \end{pmatrix}
    \label{eq:plusExpandedHamiltonian}
\end{equation}
and
\begin{equation}
    H^{(-)}=
    \begin{pmatrix}
        0 & -v_3 p^\dagger & 0 & -v_0 p \\ -v_3 p & 0 & -v_0 p^\dagger & 0 \\ 0 & -v_0 p & 0 & -\gamma_1 \\ -v_0 p^\dagger & 0 & -\gamma_1 & 0
    \end{pmatrix},
    \label{eq:minusExpandedHamiltonian}
\end{equation}
where the superscripts $(+)$ and $(-)$ refer to the valleys. Here, $p = p_x + ip_y$ is the complex representation of the two-dimensional quasi-momentum relative to the Dirac point, $v_j = 3a\gamma_j/2\hbar$ for $j=0,3$, such that $v_0\simeq10^6\ \si{m s^{-1}}$, and $v_3\simeq1.2 \times10^5\ \si{m s^{-1}}$.

\begin{figure}
    \centering
    \includegraphics[width=70mm]{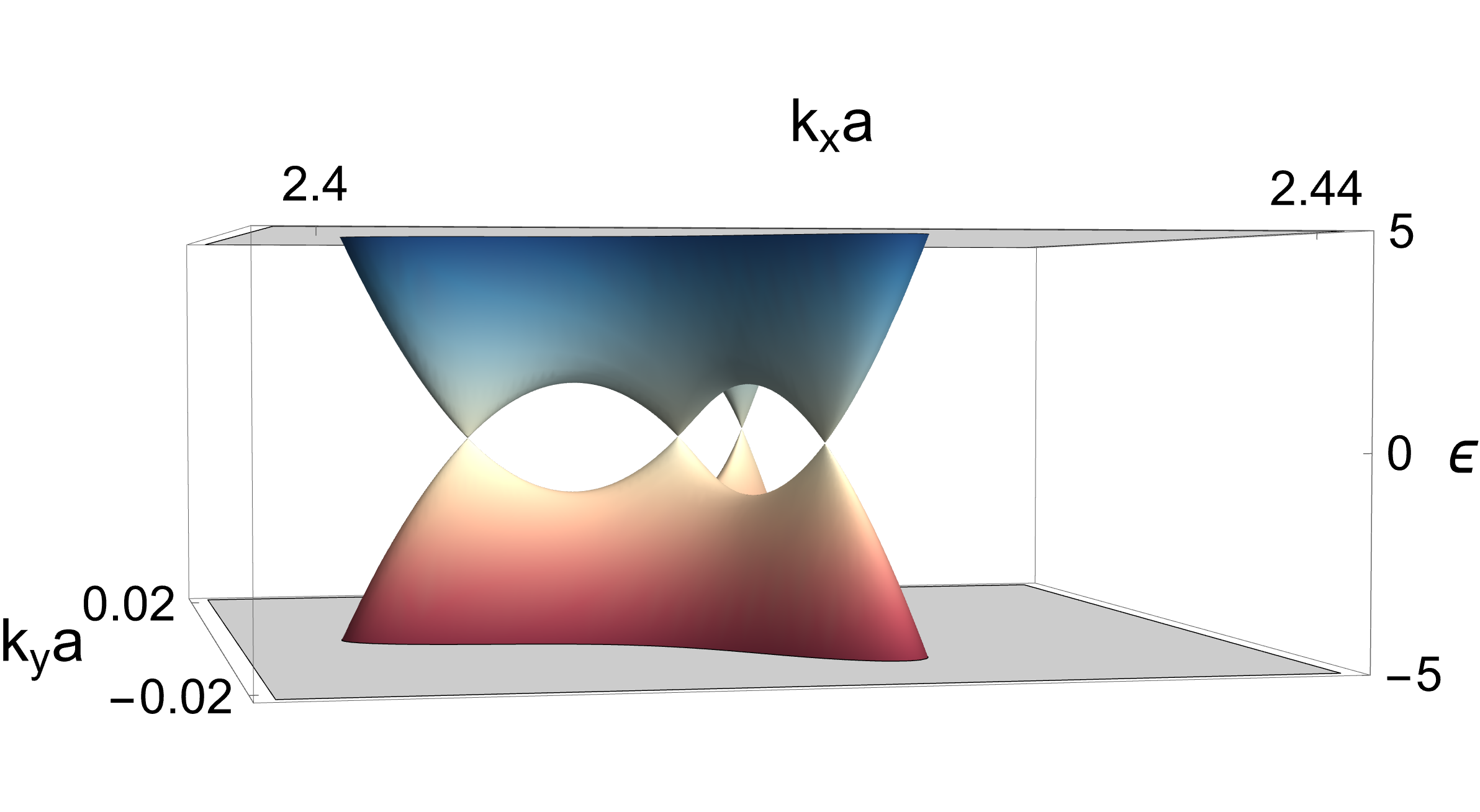}\\ \vspace{15pt}
    \includegraphics[width=80mm]{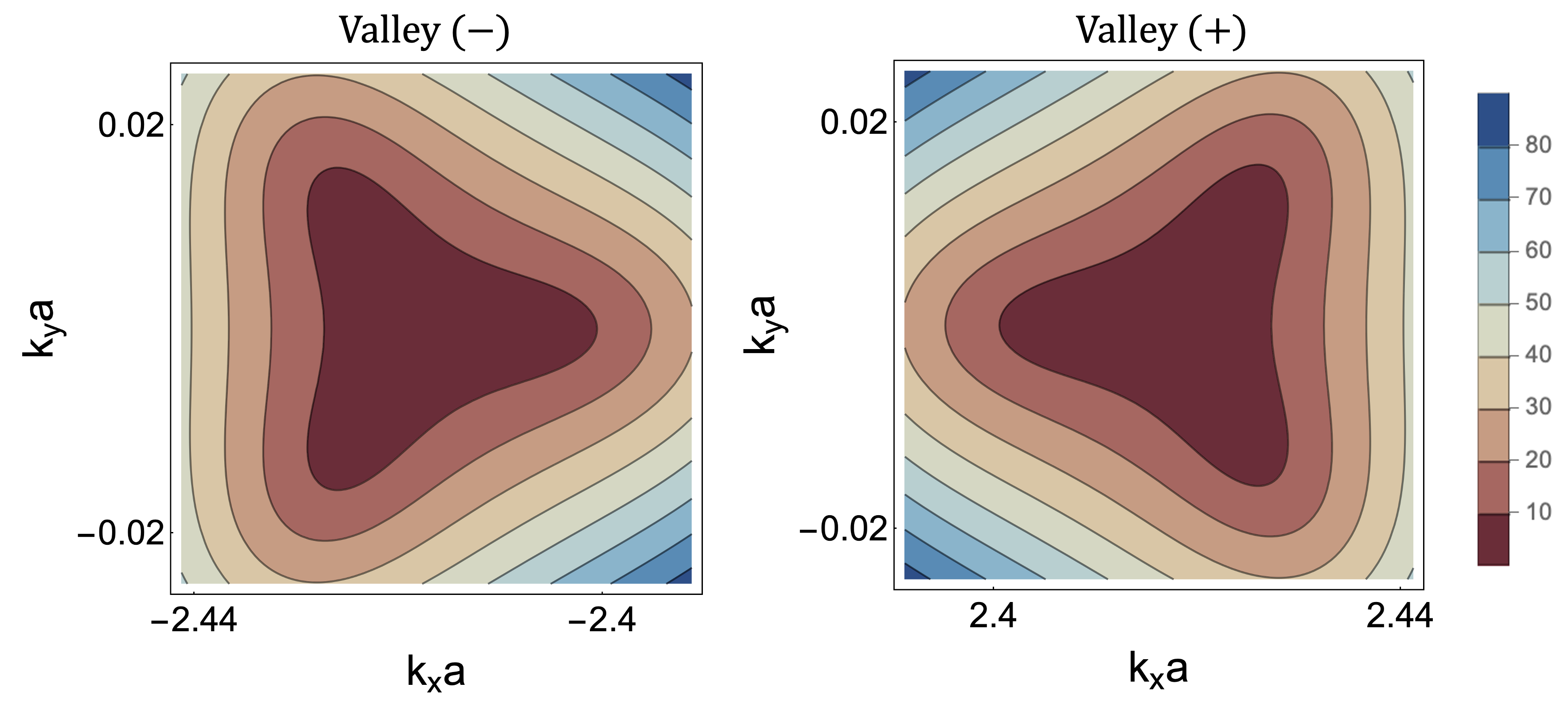}
    \caption{(Color online) Low-energy band structure of bilayer graphene ($\epsilon$ in $\si{m eV}$) originating from the 4x4 Hamiltonian in Eq. (\ref{eq:bilayerHamiltonian}). The lower panels show equienergy contours of bilayer graphene's band structure for the two nonequivalent valleys. There is a clear valley-dependent anisotropy that survives up to energies of order $\sim 100 \si{meV}$.}
    \label{fig:band_structure}
\end{figure}

An effective $2\times 2$ Hamiltonian at low energies can be derived for each valley using the Schrieffer-Wolff transformation \cite{bravyi2011schrieffer} (see Appendix for a derivation). This gives the Hamiltonian
\begin{equation}
    H_{\mathrm{eff}}^{(\xi)}=
    \begin{pmatrix}
        0 & g_{\xi}(p) \\ g_{\xi}^*(p) & 0
    \end{pmatrix},
    \label{eq:SWHamiltonian}
\end{equation}
where $\xi$ refers to the valley (either $(+)$ or $(-)$), 
\begin{equation}
    g_{+}(p) = v_3 p + \frac{v_0^2}{\gamma_1} {p^{\dagger}}^2
    \label{eq:gpPlus}
\end{equation}
and
\begin{equation}
    g_{-}(p) = -v_3 p^{\dagger} + \frac{v_0^2}{\gamma_1} p^2.
    \label{eq:gpMinus}
\end{equation} 
The terms proportional to $v_3$ are associated with trigonal warping that produces a qualitative restructuring of the low energy electronic band structure (see Fig. \ref{fig:band_structure} (upper panel)). This will have dramatic consequences on the photoexcitation properties of the system. 

The eigenenergies $E_{\pm}^{(\xi)}(p)$ and normalized eigenstates $\ket{\xi,\pm}$ of the Hamiltonian (\ref{eq:SWHamiltonian}) are
\begin{equation}
    E_{\pm}^{(\xi)}(p) = \pm |g_{\xi}(p)|
    \label{eq:eigenval1}
\end{equation}
and
\begin{equation}
    \ket{\xi,\pm} = \frac{1}{\sqrt{2}}\left(\frac{\pm g_{\xi}(p)}{|g_{\xi}(p)|},\ 1\right)^T.
    \label{eq:eigenvec1}
\end{equation}

Figure \ref{fig:band_structure} (upper panel) shows the eigenenergies (\ref{eq:eigenval1}) for the $(+)$ valley up to $5 \si{meV}$. At these low energies, the band structure is characterized by four mini-Dirac cones that merge into singly-connected equipotential domains above the Lifshitz transition energy of order $\sim 1 \si{meV}$. This valley-dependent highly anisotropic band structure survives up to energies of order $\sim 100 \si{meV}$ as shown by the equipotential lines in Fig. \ref{fig:band_structure} (lower panel). We will show in section \ref{sec:polarplots} that this phenomenon may be exploited to create valley separation of charge carriers in bilayer graphene.

A band gap can be opened in bilayer graphene by applying an electric field perpendicular to the layers \cite{zhang2009bilayer}. The low-energy properties of gapped bilayer graphene with a band gap of $E_g=2 \Delta$ can be analyzed by introducing two diagonal terms into the effective low-energy Hamiltonian (\ref{eq:SWHamiltonian}) (see the appendix for a derivation)
\begin{equation}
    H_{\Delta}^{(\xi)}=
    \begin{pmatrix}
        \tilde{\Delta}(p) & g_{\xi}(p) \\ g_{\xi}^*(p) & - \tilde{\Delta}(p)
    \end{pmatrix},
    \label{eq:ReducedGappedSWHamiltonian}
\end{equation}
where 
\begin{equation}
    \tilde{\Delta}(p)=\Delta\left(1-\frac{2v_0^2|p|^2}{\gamma_1^2}\right).
\end{equation}
The eigenenergies $E_{\Delta,\pm}^{(\xi)}(p)$ and normalized eigenstates $\ket{\Delta,\xi,\pm}$ of the Hamiltonian (\ref{eq:ReducedGappedSWHamiltonian}) are
\begin{equation}
    E_{\Delta,\pm}^{(\xi)}(p) = \pm\sqrt{\tilde{\Delta}^2(p) + |g_{\xi}(p)|^2}
    \label{eq:gappedeigenval}
\end{equation}
and
\begin{equation}
    \ket{\Delta,\xi,\pm} = \sqrt{\frac{E^{(\xi)}_{\Delta}(p)\mp\tilde{\Delta}(p)}{2E^{(\xi)}_{\Delta}(p)}}\left(\frac{g_{\xi}(p)}{\pm E_{\Delta}^{(\xi)}(p)-\tilde{\Delta}(p)},\ 1\right)^T,
    \label{eq:gappedeigenvec1}
\end{equation}
where $E^{(\xi)}_{\Delta}(p) = |E_{\Delta,\pm}^{(\xi)}(p)|$. For convenience of notation, from now on we will drop the explicit dependencies on $p$.

\begin{figure}
    \centering
    \includegraphics[width=70mm]{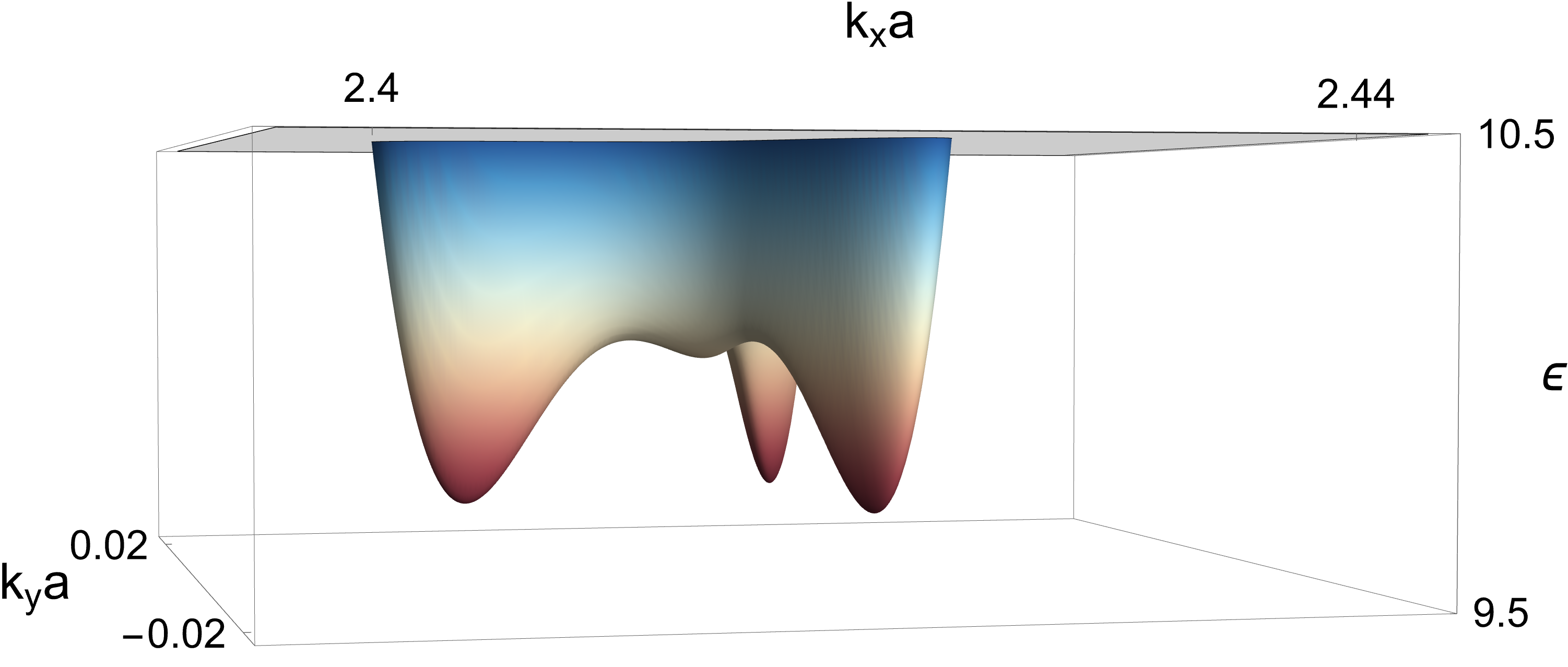}\\ \vspace{5pt}
    \includegraphics[width=70mm]{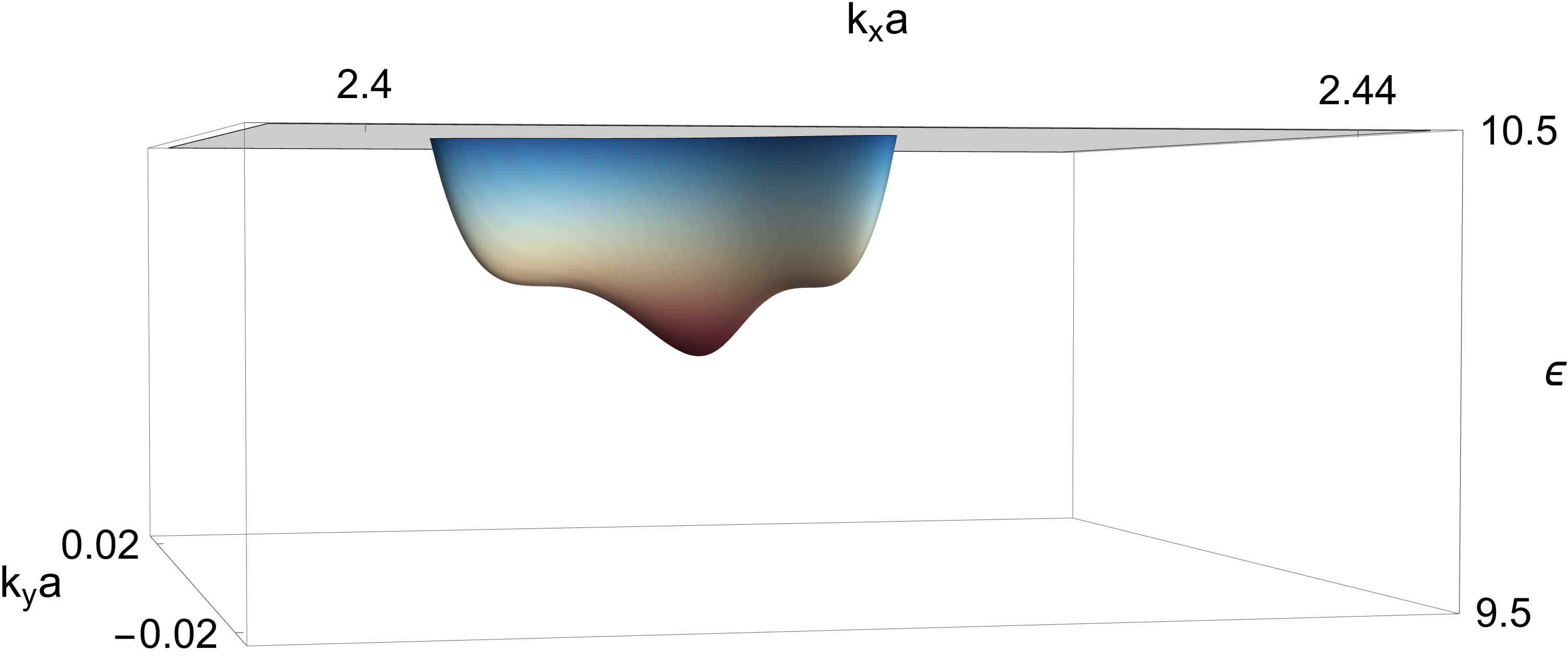}
    \caption{(Color online) Low-energy band structure (only conduction band shown) of gapped bilayer graphene ($\epsilon$ in $\si{m eV}$) with a band gap of $E_g=20 \si{meV}$. The upper panel shows the band structure stemming from the $4\times 4$ Hamiltonian in Eq.~(\ref{eq:bilayerHamiltonian}) in presence of a gap $\Delta=10 \si{meV}$. The lower panel shows the band structure including higher-order skew interlayer $A1-A2$ hopping terms, parametrised by the hopping energy $\gamma_4=0.14 \si{eV}$.}
    \label{fig:gapped_band_structure}
\end{figure}

Figure \ref{fig:gapped_band_structure} shows the low-energy conduction band of gapped bilayer graphene with a band gap of $E_g=20\si{meV}$ in the THz frequency range. Including only $\gamma_3$ and lower-order hopping terms results in a shallow central minimum surrounded by three satellite minima at lower energy, as shown in the upper panel. Including higher-order skew interlayer $A1-A2$ hopping terms, parametrized by the hopping energy $\gamma_4$, has the effect of raising the three satellite minima, as shown in the lower panel. Depending on the value of $\gamma_4$, which is not precisely known, the three satellite minima may rise above the central one. According to McCann \cite{mccann2013electronic}, $\gamma_4\simeq0.14 \si{eV}$, which indeed results in a central global minimum at the $\bm{K}_{+}/\bm{K}_{-}$ points. It should be noted that the inclusion of the $\gamma_4$ terms does not affect the main results of this work, in particular the ability to separate photo-excited carriers from different valleys and to detect them, as discussed below. 

\section{Optical Selection Rules For Bilayer Graphene}

\subsection{Gapless Bilayer Graphene}

We now derive the matrix elements for optical transitions in pristine bilayer graphene.
In the dipole approximation, the transition rate $W_{\bm{k}}$ of an electron, of wave vector $\bm{k}$, from the valence band to the conduction band is given by Fermi's golden rule:
\begin{equation}
    W_{\bm{k}}=\frac{\alpha I_e}{\nu^2}\left|\bm{e}\cdot\bra{\psi_C(\bm{k})}\hat{\bm{v}}\ket{\psi_V(\bm{k})}\right|^2\delta(E_C-E_V-h\nu),
    \label{eq:fermi}
\end{equation}
where $\ket{\psi_C(\bm{k})}$ and $\ket{\psi_V(\bm{k})}$ are the eigenstates of electrons in the conduction and valence bands, and $E_C$ and $E_V$ are the corresponding energies. Here, $\alpha=e^{2}_{}/4\pi\epsilon_{0}^{}\hbar c\simeq1/137$ is the fine-structure constant, while $\bm{\hat{v}}$, $I_e$, and $\nu$ are the velocity operator, intensity and frequency of the excitation, respectively. The vector $\bm{e}=(e_x,e_y)^T$ describes the polarization of the excitation, which we assume to propagate normal to the surface of the crystal \cite{saroka2022momentum,hartmannthesis}. Within the gradient approximation, the velocity operator is deduced from the effective $2\times 2$ Hamiltonian (\ref{eq:SWHamiltonian}) as
\begin{equation}
    \bm{\hat{v}}^{(\xi)} = \frac{\partial H^{(\xi)}_{\mathrm{eff}}}{\partial \bm{p}} = \begin{pmatrix}
        0 & \nabla_{p} g_{\xi} \\ \nabla_{p} g_{\xi}^* & 0
    \end{pmatrix},
\end{equation}
where $\bm{p} = (p_x,p_y)^T$. Thus, the optical selection rules of our system stem from the matrix element ${M^{(\xi)}} = \bm{e}\cdot \bra{\xi,+}\bm{\hat{v}}^{(\xi)}\ket{\xi,-}$ given by
\begin{equation}
    {M^{(\xi)}} = \bm{e}\cdot\frac{i}{|g_{\xi}|}\Im(g_{\xi}^* \nabla_p g_{\xi}),
    \label{eq:mel}
\end{equation}
where $\Im$ denotes the imaginary part.

In section \ref{sec:optovalleytronics}, we will use Eq. (\ref{eq:mel}) to derive the angular generation density of photoexcited carriers, and show that valley separation is achievable in bilayer graphene at low excitation energies using linearly-polarized light.

\subsection{Gapped Bilayer Graphene}
\label{sec:gappedselectionrules}

\begin{figure}
    \centering
    \includegraphics[width = 80mm]{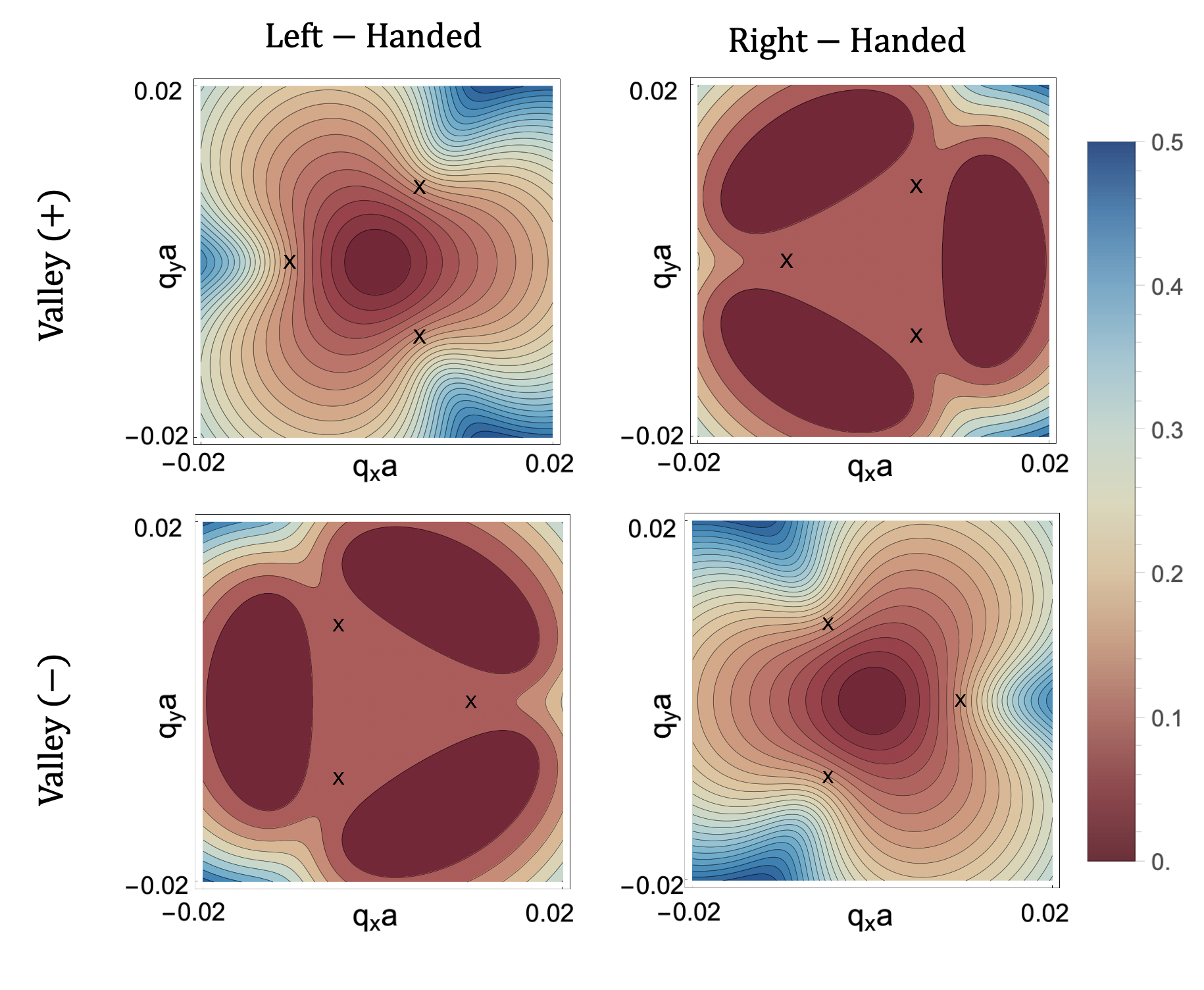}
    \caption{(Color online) Contour plots of the modulus-squared matrix element ($|M_{\Delta}^{(\xi)}|^2$) of the velocity operator (in units of $v_0^2$) between the valence and conduction bands in bilayer graphene with a band gap of $E_g=20 \si{m eV}$. The rows show the results for circularly-polarized light of left- and right- handedness for each valley. There is a clear valley-dependence on the handedness of circularly-polarized light associated with optical transitions. Here, $q_x$ and $q_y$ are the wavevector components measured from the Dirac points. The locations of the satellite minima (see Fig.\ref{fig:gapped_band_structure}) are marked by black crosses.}
    \label{fig:gapped_selection}
\end{figure}

The approach above can be generalized to deduce the optical section rules for gapped bilayer graphene as well.
The matrix element $M_{\Delta}^{(\xi)} = \bm{e}\cdot\bra{\Delta,\xi,+}\hat{\bm{v}}^{(\xi)}\ket{\Delta,\xi,-}$ for optical transitions for the Hamiltonian (\ref{eq:ReducedGappedSWHamiltonian}) is
\begin{multline}
M_{\Delta}^{(\xi)} = \frac{\bm{e}}{|g_{\xi}|}\cdot\Bigg(\frac{\tilde{\Delta}}{E_{\Delta}^{(\xi)}}\Re(g_{\xi}^*\nabla_p g_{\xi})+\\+i \Im(g_{\xi}^*\nabla_p g_{\xi}) -\frac{|g_{\xi}|^2}{E_{\Delta}^{(\xi)}}(\nabla_{p}\tilde{\Delta})\Bigg),
\label{eq:gappedmel}
\end{multline}
where $\Re$ denotes the real part. This reduces to the gapless result in Eq.~(\ref{eq:mel}) in the limit $\Delta\to 0$.

Figure \ref{fig:gapped_selection} shows contour plots of the modulus-square of the matrix element (\ref{eq:gappedmel}), for the circular polarizations $\bm{e} = \frac{1}{\sqrt{2}}(1,\pm i)^T$, with $\Delta=10\si{meV}$. The $(+)$ and $(-)$ correspond to left- and right-handed circular polarizations, respectively. There is a clear coupling between the valley index and the handedness of circularly-polarized light associated with optical transitions in gapped bilayer graphene. 

The presence of trigonal warping is directly evident in the angular dependence of the matrix element. Moreover, the linear terms in $p$ in the effective Hamiltonian, proportional to $v_3$, induce a crossover between the selection rules as a function of $|p|$. Indeed, in the $(+)$ valley (upper row in Fig.~\ref{fig:gapped_selection}) the optical transitions are dominated by the right-handed circular polarization at the $\bm{K}_{+}$ point ($|p|=0$) and by the left-handed circular polarization at larger momenta. The effect is opposite in the $(-)$ valley (lower row in Fig.~\ref{fig:gapped_selection}). Notice that this crossover would not be present if trigonal warping was neglected, as conventionally done in the literature.  

These valley-dependent selection rules can be exploited to detect the degree of valley polarization in gapped bilayer graphene by measuring the handedness of circularly-polarized light emitted across the sample. We will discuss this detection method in Sec.~\ref{subsec:device} below.  

\section{Optovalleytronics in Bilayer Graphene}
\label{sec:optovalleytronics}

\subsection{Angular Generation Density of Photoexcited Carriers in Bilayer Graphene}
\label{sec:polarplots}

\begin{figure*}
    \centering
    \includegraphics[width = 150mm]{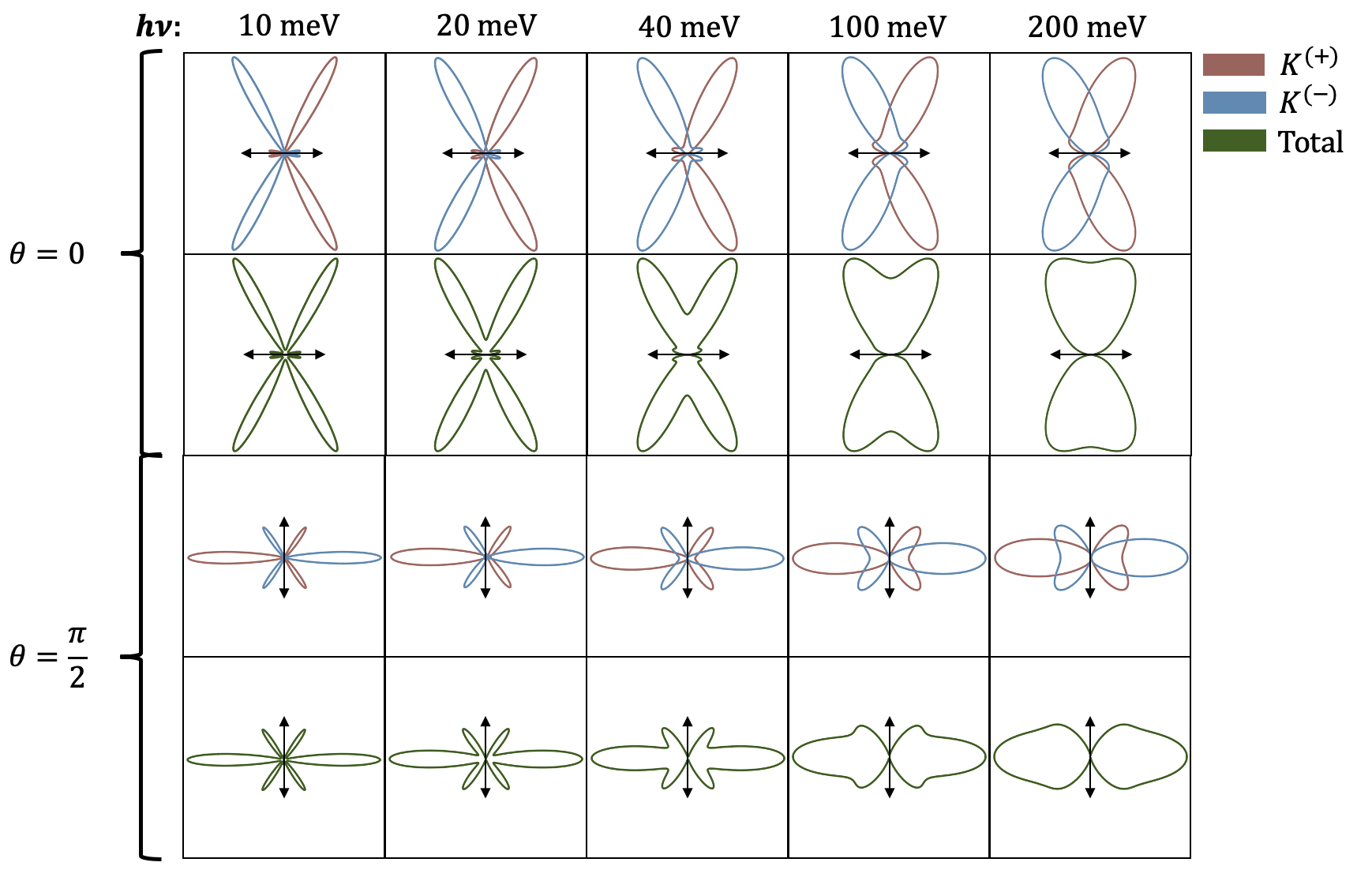}
    \caption{(Color online) Polar plots of the angular generation density $G^{(\xi)}(\varphi_{q})$ of photoexcited carriers for linearly-polarized light incident on gapless bilayer graphene ($E_g=0$) at various photon excitation energies $h\nu$. In this figure, red corresponds to the ($\xi=+$) valley, blue corresponds to the ($\xi=-$) valley, and green corresponds to their sum. The polarization angle is shown as a black, double-headed arrow. There is a clear valley-dependent spatial distribution, which may be used in valleytronic devices for the processing of quantum information. Since the anisotropy in the band structure is only present for small excitation energies, valley separation is only possible for photon energies $\epsilon_\gamma$ with $\epsilon_\gamma \lesssim 100 \si{m eV}$.}
    \label{fig:polplot}
\end{figure*}

\begin{figure}
    \centering
    \includegraphics[width=240pt]{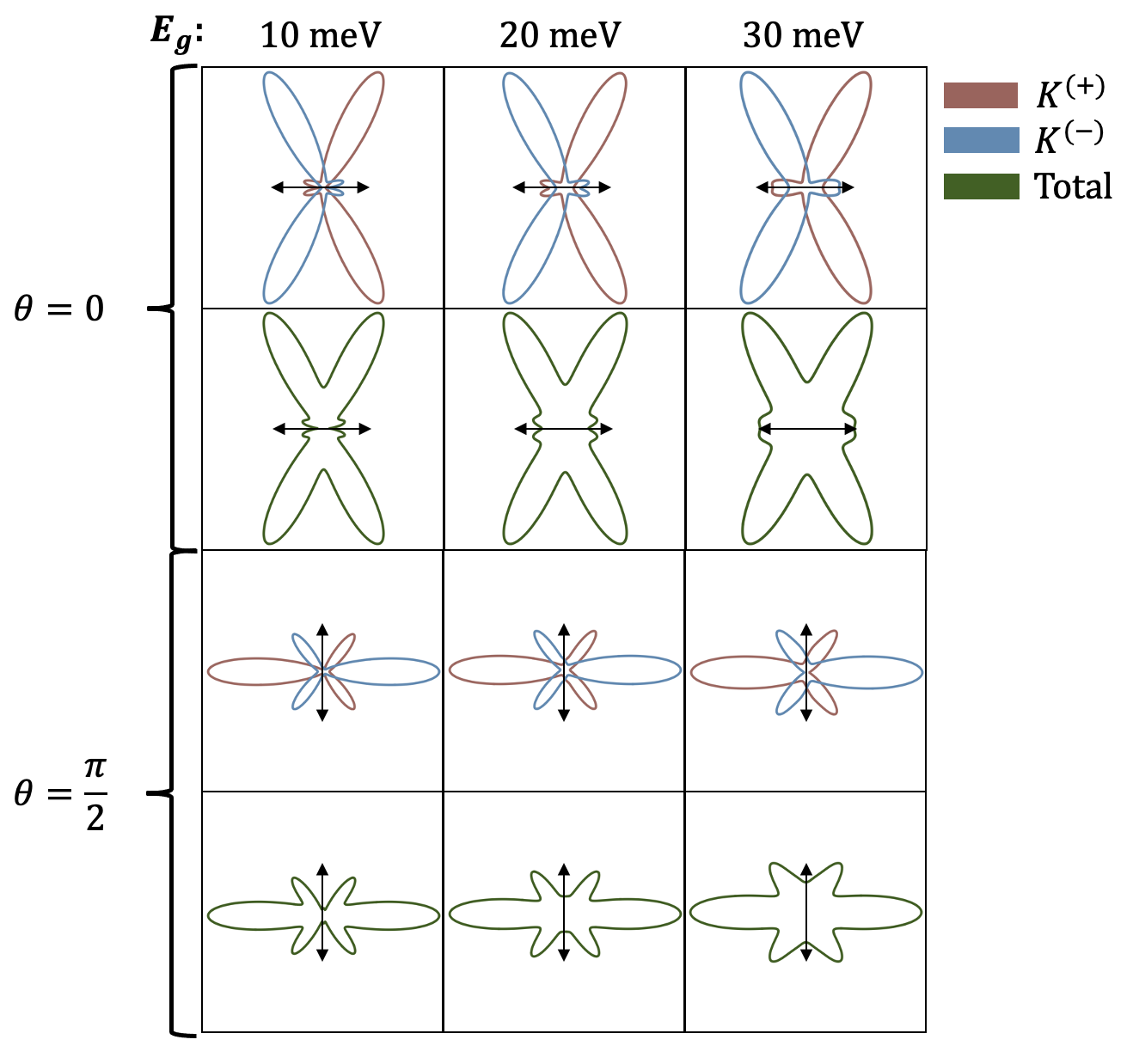}
    \caption{(Color online) Polar plots of the angular generation density $G^{(\xi)}(\varphi_{q})$ of photoexcited carriers for linearly-polarized light with a fixed photon energy $h\nu=40 \si{meV}$ in gapped bilayer graphene for varying band gaps $E_g$. In this figure, red corresponds to the ($\xi=+$) valley, blue corresponds to the ($\xi=-$) valley, and green corresponds to their sum. The polarization angle is shown as a black, double-headed arrow. Even in the presence of a gap, the valley-dependent spatial distribution observed in Fig. \ref{fig:polplot} persists.}
    \label{fig:gapped_polar}
\end{figure}

We will now use the matrix element given in Eq.~(\ref{eq:mel}) to derive the angular generation density of photoexcited carriers in the instance of photocreation in bilayer graphene.

The angular generation density $G^{(\xi)}(\varphi_{q})$ is defined such that $G^{(\xi)}(\varphi_{q})d\varphi_{q}$ gives the rate of carriers per unit area created in the angle range $\varphi_{q}$ to $\varphi_{q} + d\varphi_{q}$. For a single valley and spin, the angular generation density is given by \cite{saroka2022momentum}
\begin{equation}
    G^{(\xi)}(\varphi_{q}) = \left(\frac{1}{2\pi}\right)^2\int W_{\bm{q}}(q,\varphi_{q})qdq,
    \label{eq:anggen}
\end{equation}
where $\bm{q}=\bm{p}/\hbar$. Combining Eqs.~(\ref{eq:fermi}) and (\ref{eq:mel}) with Eq.~(\ref{eq:anggen}), and performing the integration numerically gives the angular distribution of photoexcited carriers in bilayer graphene.
Although \(G^{(\xi)}(\phi_q)\) is defined as a distribution over the wavevector angle, the real-space propagation direction of a photoexcited carrier is determined by its group velocity,
\(\mathbf{v}^{(\xi)}_g=\hbar^{-1}\nabla_{\mathbf{q}}E^{(\xi)}_+(\mathbf{q})\).
This group velocity is normal to the contours of equienergy illustrated in Fig.~\ref{fig:band_structure}. In the trigonally warped regime, the normals associated with the three lobes are approximately aligned with the lobe directions, so \(G^{(\xi)}(\phi_q)\) directly indicates the preferred directions of carrier propagation.

Figure \ref{fig:polplot} shows polar plots of $G^{(\xi)}(\varphi_{q})$ for gapless bilayer graphene for linearly-polarized light at various excitation energies, for polarization angles $\theta=0$ and $\theta=\frac{\pi}{2}$. The red and blue plots show the contribution from the $(+)$ and $(-)$ valleys, respectively, while the green plots show their sum, and the double-headed black arrow represents the polarization of the excitation.

It can be seen that at low energies, there is a large degree of valley separation in the instance of photocreation. If the light has a polarization angle of $\theta = \frac{\pi}{2}$, the charge carriers propagate away from the light spot preferentially along the positive $x$-axis for the $(-)$ valley and the negative $x$-axis for the $(+)$ valley. Notice that, due to momentum alignment phenomena~\cite{hartmannthesis,saroka2022momentum}, there are no carriers generated in the direction parallel to the polarization plane of the light. 

This valley separation persists even in the case of gapped bilayer graphene for experimentally attainable values of the band gap ($E_g=2\Delta$), and photon excitation energies above $E_g$, as shown in Fig. \ref{fig:gapped_polar}. Together with the selection rules for circularly-polarized light of gapped bilayer discussed in Sec.~\ref{sec:gappedselectionrules}, this leads to the optical valley Hall effect - the emission of two different circular polarizations from different sides of the light spot, upon linearly-polarized excitation.

It is important to stress that this asymmetry persists up to excitation energies exceeding $100\si{meV}$ where the low-energy restructuring due to trigonal warping is not as prominent. Therefore, we expect the observation of this effect to be robust against charge-density fluctuations due to local inhomogeneities. This valley separation mechanism may be used to independently manipulate charge carriers in different valleys, for potential use in future valleytronic devices. 

\subsection{Basic Optovalleytronic Devices Using Bilayer Graphene}\label{subsec:device}

\begin{figure}
    \centering
    \includegraphics[width=70mm]{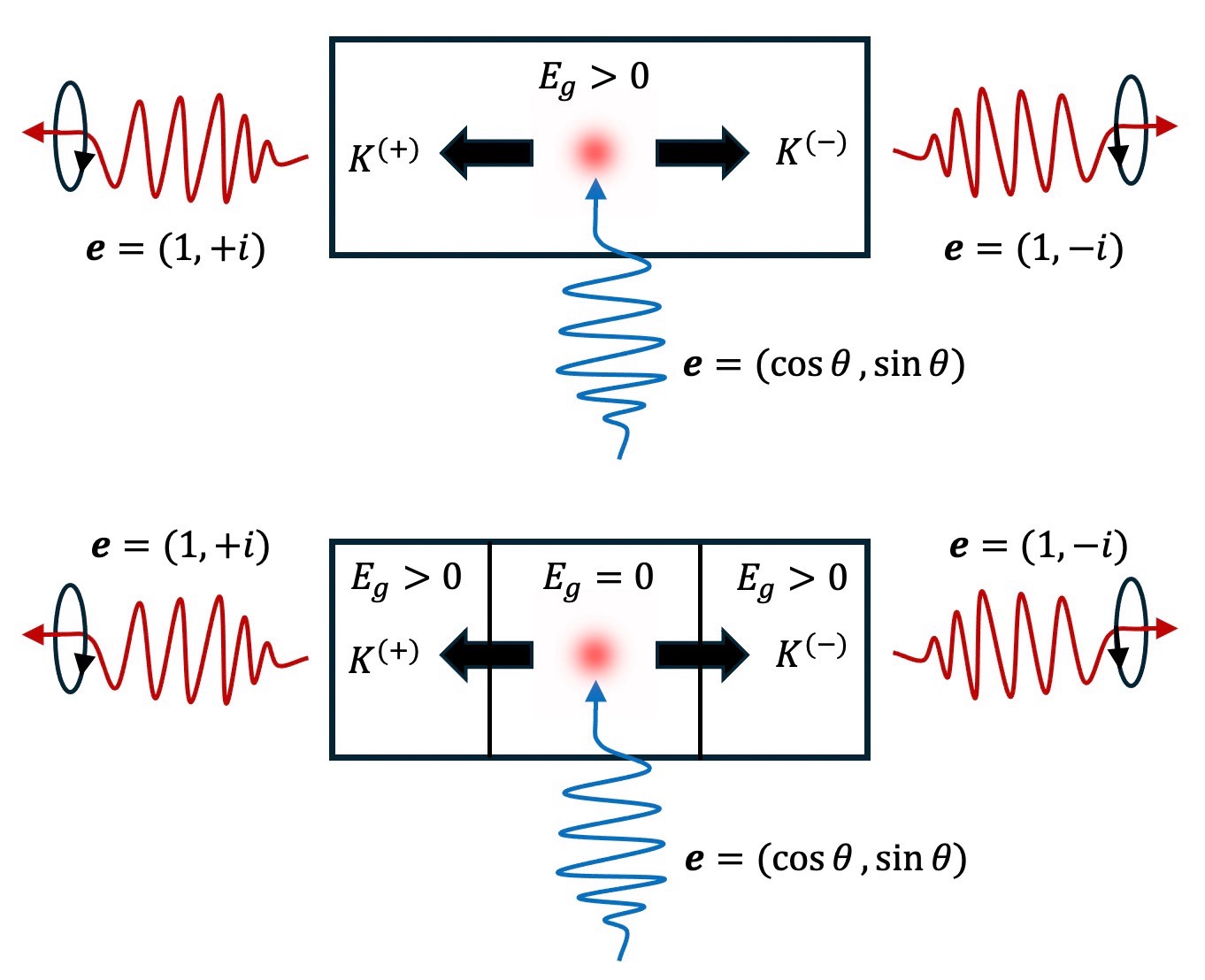}
    \caption{(Color online) Schematics of rudimentary optovalleytronic devices using bilayer graphene. (Upper panel) Uniformly gapped bilayer graphene excited above the band gap with linearly-polarized light. Charge carriers in different valleys propagate away from the light spot (shown by the red dot) because of the anisotropic band structure. Due to the optical selection rules, band-edge photoluminescence has different circular polarizations at different sides of the light spot. (Lower panel) A device with interfaces between gapless and gapped bilayer graphene. Linearly-polarized light is shone on the gapless region of the sample ($E_g = 0$). Charge carriers with different valley indices propagate into opposite gapped regions ($E_g >0$) where they emit circularly-polarized light with opposite handedness.}
    \label{fig:valleytronic_device}
\end{figure}

We now propose two rudimentary experimental setups exploiting the properties of gapless and gapped bilayer graphene derived in sections \ref{sec:gappedselectionrules} and \ref{sec:polarplots} to create and measure valley separation. The two setups are illustrated in Fig. \ref{fig:valleytronic_device}. 

In the upper panel, there is a scheme for observing the optical valley Hall effect in uniformly-gapped bilayer graphene. Carriers should be excited by linearly-polarized light (red light spot in Fig.~\ref{fig:valleytronic_device}) above the band gap resulting in the anisotropic momentum and valley-index distribution, as shown in Fig. \ref{fig:gapped_polar}, and in the consequent spatial separation of the carriers from different valleys. For low enough excitation energies (below the inter-valley phonon energy), photoexcited carriers at the two opposite sides of the light spot will relax to the band edges (at the three satellite minima indicated by crosses in Fig.~\ref{fig:gapped_selection}) retaining their valley index. The consequent edge photoluminescence will be dominated by left (right) -handed circular polarization in the $\bm{K}_{+} (\bm{K}_{-})$ valley. To observe the effect, the sample should be doped (or gate-doped) so that photoexcited carriers recombine with their equilibrium counterparts. Thus, the valley index of minority carriers is detected.

Notice that this detection approach is robust against the inclusion of higher-order hopping terms. Depending on the value of $\gamma_4$ (see Fig.~\ref{fig:gapped_band_structure}), photoexcited carriers may instead relax to the band edges at the $\bm{K}_{+} (\bm{K}_{-})$ points. In this case, they would predominantly emit right (left) -handed circularly polarized light in the $\bm{K}_{+} (\bm{K}_{-})$ valley. Importantly, the degree of circular polarization from opposite sides of the light spot would still be opposite for the two nonequivalent valleys, allowing for the detection of the valley polarization.

In the lower panel of Fig.~\ref{fig:valleytronic_device}, a central region of gapless bilayer graphene is surrounded by two gapped regions (induced by means of top- and back-gate voltages). In contrast to the previous setup, this device benefits from a higher mobility in the central gapless region. In this device, linearly-polarized light is shone onto gapless bilayer graphene, resulting in high-mobility charge carriers of different valley-index propagating into the two gapped regions. In any realistic setup, the gate-induced gap changes smoothly over length scales much larger than the lattice spacing, thereby preventing inter-valley mixing at the gapless-gapped interface. Circularly-polarized light of opposite handedness should then be emitted at the two different gapped regions. In both setups, the incident photon energy must be below the inter-valley phonon energy of $0.16 \si{eV}$ to avoid inter-valley scattering.

We emphasize that our proposed setups: (i) exploit bulk properties of bilayer graphene, and their functionality does not involve any edge effects that are known to deteriorate the valley index in conventional transport experiments; (ii) are based on conventional bilayer devices and excitation frequencies that have been available in the literature for over a decade \cite{zhang2009bilayer}; (iii) operate in the elusive terahertz regime, a frequency range critical for emerging technologies, including 6G communications, security screening, and advanced testing in pharmaceutical and biomedical applications.

\vspace{62.5pt}

\section{Conclusions}
We find that, by exploiting the highly anisotropic nature of the electronic band structure of gapless and gapped bilayer graphene at low energies ($E \lesssim 100\si{meV}$), spatial separation of charge carriers in different valleys is possible using optical excitation with low-frequency photons. This effect can be enhanced via momentum alignment induced by linearly-polarized light. Importantly, in this low-energy regime, inter-valley electron-phonon scattering is suppressed, thereby protecting the valley index. This is in stark contrast to the band structure of monolayer graphene, which only begins to deviate significantly from the isotropic conical dispersion near the Dirac points well above the inter-valley phonon energy of $0.16\si{eV}$. Additionally, gapped bilayer graphene exhibits valley-dependent optical selection rules, which may be used to measure the degree of valley polarization of the spatially separated charge carriers. We propose two realistic experimental setups that exploit these effects in gapless and gapped bilayer graphene as a basis for future optovalleytronic devices.
Recently, it has been experimentally shown \cite{Ensslin2024} that valley states in bilayer graphene are long-lived, with lifetimes an order of magnitude longer than those of conventional spin states.
This makes bilayer graphene a highly promising platform for emerging quantum-optovalleytronics applications.  

\begin{acknowledgments}
This work was supported by the UK EPSRC grant No. EP/Y021339/1 and by the EU HORIZON-MSCA-SE project HERMES (Grant agreement ID: 101236439). E.M. acknowledges insightful discussions with Mauro and Roberta Bucanieri. 
For the purpose of open access, the authors have applied a CC BY public copyright license to any Author Accepted Manuscript version arising from this submission.
\end{acknowledgments}

\appendix*
\section{Low Energy Effective Hamiltonian of Bilayer Graphene}
\label{sec:appendices}

In this Appendix, we derive the low energy effective Hamiltonian of bilayer graphene in Eq. (\ref{eq:SWHamiltonian}), starting from the Hamiltonian (\ref{eq:bilayerHamiltonian}).
We use the Schrieffer-Wolff (SW) transformation \cite{bravyi2011schrieffer} to decouple the Hamiltonian into low- and high-energy subspaces. For our purposes with bilayer graphene, this corresponds to deriving a Hamiltonian for the two bands shown in Fig. \ref{fig:band_structure} (upper panel) which touch close to zero energy, whilst neglecting the other two that correspond to higher energy excitations of order $\pm \gamma_1$. For our analysis, the relevant low-energy subspace is therefore spanned by the $A1$ and $B2$ sites, since these are the non-dimer sublattices: unlike the $A2$ and $B1$ sites, they are not directly coupled by the strong interlayer hopping $\gamma_1$, and hence are not split away from zero energy into the high-energy bonding and antibonding dimer states \cite{Mariani2012}.

In the SW transformation, we begin with a Hamiltonian of the form
\begin{equation}
    H = H_0 + V, 
    \label{eq:SWForm}
\end{equation}
where $H_0$ is block diagonal, and $V$ contains terms that mix the low- and high-energy subspaces.

We now perform a unitary transformation, using the unitary operator $U = e^{-S}$, with the generator $S$, to obtain the Hamiltonian $H'$, given by
\begin{equation}
    H' = U^{\dagger}HU = e^S H e^{-S},
\end{equation}
with $S^{\dagger} = - S$, chosen so that $H'$ is block diagonal to the desired order. Expanding $U^\dagger$ to second-order in $S$ gives
\begin{equation}
    U^{\dagger} = e^S = \sum_{n=0}^{\infty} \frac{S^n}{n!}\simeq (1 + S + \frac{1}{2}S^2).
\end{equation}
Therefore,
\begin{equation}
\begin{split}
        H' & \simeq (1 + S + \frac{1}{2}S^2) H (1 - S + \frac{1}{2}S^2)\\
        & \simeq H + [S,H] + \frac{1}{2}[S,[S,H]],
\end{split}
\label{eq:SWWorking1}
\end{equation}
which is a special case of the Baker-Campbell-Hausdorff formula.

At leading order, we then choose $S$ to cancel the off-diagonal coupling i.e. $V_{\mathrm{off}} + [S,H_0] = 0$. Substituting this condition into Eq. (\ref{eq:SWWorking1}) and projecting onto the low-energy subspace with the projector $P$ gives the low-energy effective Hamiltonian
\begin{equation}
    H'_{(P)} = PH_0P + PVP + \frac{1}{2} P[S,V_{\mathrm{off}}]P.
    \label{eq:SWEquation}
\end{equation}

In second quantisation, the Hamiltonian in Eq. (\ref{eq:bilayerHamiltonian}), supplemented by a layer potential asymmetry that opens a band gap of \(2\Delta\), can be written in the form of Eq. (\ref{eq:SWForm}) as
\begin{multline}
    H_0 = \sum_{\bm{k}} \left[-\gamma_3 (f(\bm{k}) A1_{\bm{k}}^\dagger B2_{\bm{k}} + f^*(\bm{k}) B2_{\bm{k}}^\dagger A1_{\bm{k}})\ + \right.\\ \left. -\gamma_1(A2_{\bm{k}}^\dagger B1_{\bm{k}} +B1_{\bm{k}}^\dagger A2_{\bm{k}}) + \right.\\ \left.
    + \Delta (A1_{\bm{k}}^\dagger A1_{\bm{k}} + B1_{\bm{k}}^\dagger B1_{\bm{k}})+ \right.\\ \left.
    -\Delta(A2_{\bm{k}}^\dagger A2_{\bm{k}}+B2_{\bm{k}}^\dagger B2_{\bm{k}})\right],
\end{multline}
\begin{multline}
    V = \sum_{\bm{k}} \left[-\gamma_0 \left(f^*(\bm{k}) A1_{\bm{k}}^\dagger B1_{\bm{k}} + f(\bm{k}) B2_{\bm{k}}^\dagger A2_{\bm{k}}\ + \right.\right.\\ \left.\left. + f^*(\bm{k}) A2_{\bm{k}}^\dagger B2_{\bm{k}} + f(\bm{k}) B1_{\bm{k}}^\dagger A1_{\bm{k}} \right)\right],
\end{multline}
where we have introduced the creation (annihilation) operators $A1_{\bm{k}}^\dagger, A2_{\bm{k}}^\dagger, B1_{\bm{k}}^\dagger, B2_{\bm{k}}^\dagger$ ($A1_{\bm{k}}, A2_{\bm{k}}, B1_{\bm{k}}, B2_{\bm{k}}$) which add (remove) electrons on the A/B site in the 1st/2nd layer, respectively. These creation and annihilation operators fulfill the anticommutation relations
\begin{equation}
\begin{split}
    \{ Al_{\bm{k}}, Am_{\bm{k}}^\dagger \} & = \{ Bl_{\bm{k}}, Bm_{\bm{k}}^\dagger \} = \delta_{lm}, \\
    \{ Al_{\bm{k}}, Bm_{\bm{k}}^\dagger \} & = \{ Al_{\bm{k}}, Am_{\bm{k}} \} = \{ Bl_{\bm{k}}, Bm_{\bm{k}} \} =0,
\end{split}
\end{equation}
where $\delta_{lm}$ is the Kronecker delta. 

After finding the generator S which satisfies the condition $V + [S,H_0] = 0$, the commutator $[S,V]$ can be computed and substituted into Eq. (\ref{eq:SWEquation}) giving $H'$. Keeping only those terms in the low-energy subspace A1-B2 we obtain the low-energy effective Hamiltonian
\begin{multline}
    H_{\mathrm{eff}} = \sum_{\bm{k}} \left[ A1_{\bm{k}}^\dagger B2_{\bm{k}}\left(-\gamma_3 f(\bm{k})+\frac{\gamma_0^2}{\gamma_1}{(f^*(\bm{k}))}^2\right)\ + \right. \\ \left. +B2_{\bm{k}}^\dagger A1_{\bm{k}}\left(-\gamma_3 f^*(\bm{k}) + \frac{\gamma_0^2}{\gamma_1}(f(\bm{k}))^2\right)+ \right. \\ \left. +\Delta\left(1-\frac{2\gamma_0^2}{\gamma_1^2}|f|^2\right)\left(A1_{\bm{k}}^\dagger A1_{\bm{k}}-B2_{\bm{k}}^\dagger B2_{\bm{k}}\right)
    \right],
\end{multline}
where terms beyond quadratic order in $f(\bm{k})$ have been neglected.

Taylor expanding at the Dirac point $\bm{K_{+}}$ yields the Hamiltonian in matrix form
\begin{equation}
    H_{\mathrm{eff}}^{(+)}=
    \begin{pmatrix}
        \Delta\left(1-\frac{2v_0^2|p|^2}{\gamma_1^2}\right) & v_3 p + \frac{v_0^2}{\gamma_1}{p^\dagger}^2 \\ v_3 p^\dagger + \frac{v_0^2}{\gamma_1}p^2 & -\Delta\left(1-\frac{2v_0^2|p|^2}{\gamma_1^2}\right)
    \end{pmatrix},
\end{equation}
as in Eq. (\ref{eq:SWHamiltonian}) (with $\Delta =0$) and Eq. (\ref{eq:ReducedGappedSWHamiltonian}).

\end{document}